\documentclass[12pt]{iopart}

\usepackage{iopams}   %
\usepackage{amsfonts} %
\usepackage{amssymb}  %
\usepackage{longtable}%
\usepackage{latexsym} %
\usepackage{amsthm}   %
\newcommand{\R}{\mathcal{R}}         %
\newcommand{\F}{\mathcal{F}}         %
\newcommand{\Q}{\mathcal{Q}}         %
\eqnobysec

\begin{document}

\title[Inverse contour representation as a solution $\ldots$]{Inverse contour representation as a solution of the rotating Morse potential}

\author{S A Yahiaoui and M Bentaiba}

\address{LPTHIRM, D\'epartement de Physique, Facult\'e des Sciences, Universit\'e Sa\^ad DAHLAB de Blida, B.P. 270 Route de Soum\^aa, 09000 Blida, Algeria}
\ead{s\_yahiaoui@univ-blida.dz and bentaiba@univ-blida.dz}
\begin{abstract}
A new way for obtaining the bound-states for arbitrary non zero $l$-states of the rotating Morse potential is presented. We show that by making use of the inverse contour representation, which is based on a knowledge of the integral representation of Euler's beta function, the radial wave-function for the rotating Morse potential as well as their energy eigenvalues are deduced. The results obtained are compared with the findings in the literature and it is found that are good agreement with those deduced by others methods.
\end{abstract}

\pacs{02.30.Mv; 02.30.Uu; 03.65.Ge; 34.20.Gj}
\vspace{2pc}
\submitto{\PS}
\maketitle

%%%%%%%%%%%%%%%%%%%%%%%
\section{Introduction}%
%%%%%%%%%%%%%%%%%%%%%%%

\indent The Morse potential \cite{1} is one of the simplest examples of the Natanzon potentials \cite{2} which has finite number of bound-states. It was introduced by P. M. Morse in 1929 as a model for describing the vibrating energy of a diatomic molecule and takes the form
\begin{eqnarray}\label{1}
V_{\rm M.}(x)=V_0 \left(\rme^{-2\alpha x}-2\,\rme^{-\alpha x}\right),
\end{eqnarray}
where $x=(r-r_0)/r_0$ and $\alpha=a\,r_0$. Here $V_0$ represents the dissociation energy of molecule, $r_0$ is a bound length and $a$ is a parameter to control the width of the potential.\\
\indent However, the radial Schr\"odinger equation of this potential for arbitrary non-zero $l$-states $(l\neq0)$ cannot be solved exactly unless by using various approximate schemes. Among them, the most convenient one is the Pekeris approximation \cite{3} where the basic idea of the method is so that the effective $l$-dependent potential keeps the same form as the potential with $l=0$ ($s$-state). Using this approximation, the rotating Morse potential has been solved using some useful methods and approaches including the supersymmetry (SUSY) \cite{4}, the variational approach \cite{5}, the Nikifovov-Uvanov method (NU) \cite{6}, the shifted $1/N$-expansion (SNE) \cite{7}, the modified shifted $1/N$-expansion (MSNE) \cite{8}, the asymptotic iteration method (AIM) \cite{9}, and exact quantization rule (EQR) \cite{10}.\\
\indent In the present paper, we sought to tackle the related rotating Morse bound-states problem differently from other methods approached before by performing the inverse contour representation (ICR) \cite{11} in the framework of the contour Laplace transforms, in order to solve the radial Schr\"odinger equation. The original statement for the use of ICR approach demanded a simply knowledge of the integral representation of the Euler beta function based on Cauchy's integral formula. This last restriction may be relaxed by a creation of a barrier, a cut line, joining the points $0$ and $1$, which represent the finite limit integrals of the beta function. We will see that ICR approach is conceptually simple and can be extended to others potentials.\\
\indent This work is organized as follows. In section 2 the contour Laplace transforms of the radial Schr\"odinger equation with the rotating Morse potential is outlined within Pekeris's approximation. The contour of integration $C$ is chosen so that the wave-function vanishes at the end points. Section 3 is devoted for obtaining the exact bound-states for the rotating Morse potential using the properties of ICR. In order to verify the effectiveness of our method, the energy spectra are calculated numerically for different diatomic molecules and compare with those obtained by others methods. In the last section, we do our final conclusion.

%%%%%%%%%%%%%%%%%%%%%%%%%%%%%%%%%%%%%%%%%%%%%%%%%%%%%%%%%%%%
\section{Vanishing boundary condition on the wave-function}%
%%%%%%%%%%%%%%%%%%%%%%%%%%%%%%%%%%%%%%%%%%%%%%%%%%%%%%%%%%%%

\indent The radial Schr\"odinger equation for the rotating Morse potential can be expressed as
\begin{eqnarray}\label{2}
\Bigg(\frac{\rmd^2}{\rmd r^2}+\frac{2\,m}{\hbar^2}\left[E_{n,l}-V_{\rm eff.}(r)\right]\Bigg)\R_{n,l}(r)=0,
\end{eqnarray}
where $m$ is the reduced mass of the molecule and $V_{\rm eff.}(r)$ is the effective potential given by
\begin{eqnarray}\label{3}
V_{\rm eff.}(r)&=&V_l(r)+V_{\rm M.}(r)\nonumber\\
&=&\frac{l(l+1)\hbar^2}{2\,mr^2}+V_0 \left(\rme^{-2\alpha x}-2\,\rme^{-\alpha x}\right).
\end{eqnarray}
\indent It is well-known that \eref{2} associated to the effective potential \eref{3} cannot be solved analytically for an arbitrary $l\neq0$ due to the centrifugal term. In order to obtain approximate analytical solution, we most deal approximately with the centrifugal term using the Pekeris approximation \cite{9,10} to transform the $l$-states rotating Morse potential to the $s$-state one.\\
\indent Then the centrifugal term in \eref{3} can be approximated as \cite{9,10}
\begin{eqnarray}\label{4}
\frac{l(l+1)\hbar^2}{2\,mr^2}\approx\gamma\left(a_0+a_1\,\rme^{-\alpha x}+a_2\,\rme^{-2\alpha x}\right),
\end{eqnarray}
and expanding the right-hand side of \eref{4} around $x=0$, we obtain the following parameters
\begin{eqnarray}\label{5}
\gamma=\frac{l(l+1)\hbar^2}{2\,mr_{0}^2},\quad a_0=1-\frac{3}{\alpha}+\frac{3}{\alpha^2},\quad
a_1=\frac{4}{\alpha}-\frac{6}{\alpha^2},\quad a_2=-\frac{1}{\alpha}+\frac{3}{\alpha^2}.
\end{eqnarray}
\indent Substituting \eref{3} and \eref{4} into \eref{2} the one-dimensional Schr\"odinger equation for the new rotating Morse potential $\widetilde{V}_{\rm eff.}(r)$, with respect of the variable $x$, is given by
\begin{eqnarray}\label{6}
\left(\frac{\rmd^2}{\rmd x^2}-\beta^2_2\,\rme^{-2\alpha x}+\beta^2_1\,\rme^{-\alpha x}-\epsilon^2\right)\R_{n,l}(x)=0,
\end{eqnarray}
where
\begin{eqnarray}\label{7}
\beta^2_2=\Lambda^2_0\left(V_0+\gamma a_2\right),\quad
\beta^2_1=\Lambda^2_0\left(2V_0-\gamma a_1\right),\quad
\epsilon^2=-\Lambda^2_0\left(E_{n,l}-\gamma a_0\right),
\end{eqnarray}
with $\Lambda_0=r_0\sqrt{2\,m}/\hbar$. Invoking the change in variable as well as in wave-function as
\begin{eqnarray}\label{8}
\xi=\rme^{-\alpha x}\qquad\textrm{and}\qquad\R_{n,l}(x)=\xi^\kappa \F(\xi),
\end{eqnarray}
and letting $\kappa=\epsilon/\alpha$ allow to reduce \eref{6} into
\begin{eqnarray}\label{9}
\left[\xi\frac{\rmd^2}{\rmd\xi^2}+\left(\frac{2\,\epsilon}{\alpha}+1\right)\frac{\rmd}{\rmd\xi}
-\left(\frac{\beta_2}{\alpha}\right)^2\xi+\left(\frac{\beta_1}{\alpha}\right)^2\right]\F(\xi)=0,
\end{eqnarray}
where \eref{9} accepts a regular point at $\xi=0$ and an irregular one at $\xi=\infty$.\\
\indent Now let us consider the function $\F(\xi)$ expressed in \eref{9} as an integral of the general form \cite{12}
\begin{eqnarray}\label{10}
\F_{p,q}(\xi)=\oint_{C}\Q_{p,q}(t)\,\rme^{\xi t}\,\rmd t,
\end{eqnarray}
where $\Q_{p,q}(t)$ is an unknown function to be determined.\\
\indent In \eref{10}, $C$ is the contour of integration in the real axis used to impose the vanishing boundary condition for the eigenvalue problem of \eref{9} and which does not depend on variable $\xi$. The integral \eref{10} is often called the contour Laplace transforms.\\
\indent Then by applying the $n^{\textrm{th}}$-derivative of $\F_{p,q}(\xi)$ with respect of $\xi$ and multiplying the resulting equation, \emph{i.e.} \eref{11}, by $\xi$ and performing derivation by parts, we obtain
\begin{eqnarray}\label{11}
\F_{p,q}^{(n)}(\xi)=\oint_{C}t^n\Q_{p,q}(t)\,\rme^{\xi t}\,\rmd t,
\end{eqnarray}
\begin{eqnarray}\label{12}
\xi\F_{p,q}^{(n)}(\xi)=\Big\{t^n\Q_{p,q}(t)\,\rme^{\xi t}\Big\}_{C}-\oint_{C}\frac{\rmd\left[t^n\Q_{p,q}(t)\right]}{\rmd t}\,\rme^{\xi t}\,\rmd t,
\end{eqnarray}
where the symbol $\big\{Y(t;\xi)\big\}_C$ denotes the increase of the function $Y(t;\xi)$ when $t$ describes the contour $C$. Substituting \eref{10}, \eref{11} and \eref{12} into \eref{9}, we have
\begin{eqnarray}\label{13}
\Bigg\{\Q_{p,q}(t)\left(t^2-\frac{\beta^2_2}{\alpha^2}\right)\rme^{\xi t}\Bigg\}_C-\nonumber\\
\oint_C\Bigg\{\left(t^2-\frac{\beta^2_2}{\alpha^2}\right)\Q'_{p,q}(t)+
\left[\left(1-\frac{2\epsilon}{\alpha}\right)t-\frac{\beta^2_1}{\alpha^2}\right]\Q_{p,q}(t)\Bigg\}
\,\rme^{\xi t}\,\rmd t=0,
\end{eqnarray}
where the terms in \eref{13} must vanish individually. Indeed the contour of integration $C$ must be chosen so that the first term in \eref{13} vanishes as well as the integrand at the end points; these considerations lead to solve the differential equation on $\Q_{p,q}(t)$ and we obtain
\begin{eqnarray}\label{14}
\Q_{p,q}(t)=\mathcal{N}_{p,q}\left(\frac{\beta_2}{\alpha}-t\right)^{q-1}
\left(\frac{\beta_2}{\alpha}+t\right)^{p-1},
\end{eqnarray}
where $\mathcal{N}_{p,q}$ is the reduced normalization constant. Here the parameters $p$ and $q$ are not integer and defined positive
\begin{eqnarray}\label{15}
p=\frac{\epsilon}{\alpha}+\frac{1}{2}-\frac{\beta^2_1}{2\,\alpha\beta_2},\qquad
q=\frac{\epsilon}{\alpha}+\frac{1}{2}+\frac{\beta^2_1}{2\,\alpha\beta_2}.
\end{eqnarray}
\indent Then inserting \eref{14} into \eref{10}, we obtain the solution of our problem which can be recast as
\begin{eqnarray}\label{16}
\F_{p,q}(\xi)=\mathcal{N}_{p,q}\oint_C\left(\frac{\beta_2}{\alpha}-t\right)^{q-1}
\left(\frac{\beta_2}{\alpha}+t\right)^{p-1}\rme^{\xi t}\,\rmd t,
\end{eqnarray}
and as a consequence the contour $C$, by virtue of \eref{13} and \eref{14}, must verifies the condition
\begin{eqnarray}\label{17}
\Bigg\{\left(\frac{\beta_2}{\alpha}-t\right)^{q}\left(\frac{\beta_2}{\alpha}+t\right)^{p}\rme^{\xi t}\Bigg\}_C\equiv0.
\end{eqnarray}
\indent Substituting the new change of variable $t=\beta_2(2\,z-1)/\alpha$ into \eref{16} we found that the integrand has a branch points at $z_0=0$ and $z_1=1$ and then the product $z^p\left(1-z\right)^q$ in \eref{17} will vanish for $z_0$ and $z_1$. The integrand into \eref{16} is therefore single-valued for the contour encircling both branch points; in other words, taking the line segment joining $z_0=0$ and $z_1=1$ as a cut line.\\
\indent Using the substitution $t=\beta_2(2\,z-1)/\alpha$, the function $\F_{p,q}(\xi)$ can be rewritten as
\begin{eqnarray}\label{18}
\F_{p,q}(\xi)=\rme^{-\frac{\beta_2}{\alpha}\,\xi}f_{p,q}(\xi),
\end{eqnarray}
and \eref{16} becomes
\begin{eqnarray}\label{19}
f_{p,q}(\xi)=\mathcal{N}_{p,q}\oint_C z^{p-1}\left(1-z\right)^{q-1}\rme^{\lambda\xi z}\rmd z,
\end{eqnarray}
with $\lambda=2\beta_2/\alpha$. Now \eref{19} is in a suitable form for applying the inverse contour representation as reviewed in the next section.

%%%%%%%%%%%%%%%%%%%%%%%%%%%%%%%%%%%%%%%%%%%%%%%%%%%%%%%%%%%%%%%%%%%%
\section{Inverse contour representation and bound-states solutions}%
%%%%%%%%%%%%%%%%%%%%%%%%%%%%%%%%%%%%%%%%%%%%%%%%%%%%%%%%%%%%%%%%%%%%

\indent The inverse contour representation \cite{11} is an approach which may be used to advantage in handling differential equations by means of contour integration.\\
\indent To illustrate the technique, let us derive a familiar expression of elementary functions used. Let $f(z)$ and $g(z)$ be analytic functions of $z$ on a closed contour $C$, then the last couple of functions are related to \eref{19} by an expression of the following form
\begin{eqnarray}\label{20}
f_{p,q}(\xi)&=&\frac{1}{2\,\pi\rmi}\oint_C\left(\frac{z}{z-\xi}\right)^n g_{p,q}(z)\,\rmd z,
\end{eqnarray}
and
\begin{eqnarray}\label{21}
g_{p,q}(z)&=&\int_0^1\left(1-x\right)^{n-1}f'_{p,q}(zx)\,\rmd x,\label{21}
\end{eqnarray}
where $n$ is a positive integer and $z^n/(z-\xi)^n$ is an analytic function provided with a cut line along the $\gamma$-axis joining the point $0$ and $z$. \eref{20} and \eref{21} are called direct and inverse transforms, respectively.\\
\indent We proceed here below to prove that these latter constitute a complete set of transformations; this step of evaluation can be performed consistently with the help of residues at the pole $z=\infty$ (or $u=0$) of order $k+1$, where $z=1/u$.
\begin{proof}
We shall attempt to demonstrate this assertion by substituting in \eref{20} and \eref{21} a power series with undetermined coefficients $b_k$ and $c_k$, i.e.
\begin{eqnarray}\label{22}
f_{p,q}(\xi)=\sum_{k=0}^\infty b_k\xi^k\qquad \textrm{and}\qquad g_{p,q}(\xi)=\sum_{k=0}^\infty c_k\xi^k.
\end{eqnarray}
\indent To do this inserting \eref{21} and the first summation of \eref{22} into \eref{20} and interchanging the order of integration by integrating with respect of a variable $x$ leading to express the Euler beta function, and comparing the final result with the first summation of \eref{22} we get an useful and interesting formula
\begin{eqnarray}\label{23}
\xi^k=\frac{k}{2\pi\rmi}\,B(k,n)\oint_C\frac{z^{n+k-1}}{\left(z-\xi\right)^n}\,\rmd z,
\end{eqnarray}
where $B(k,n)=\Gamma(k)\Gamma(n)/\Gamma(k+n)$ is Euler's beta function \cite{13}. The useful way to perform this integral is to use the calculus of residues; to this end the change of variable $z=1/u$ transforms the contour integral in \eref{23} into
\begin{eqnarray}\label{24}
-\frac{1}{2\pi\rmi}\oint_C\frac{\left(1-\xi u\right)^{-n}}{u^{k+1}}\,\rmd u,
\end{eqnarray}
which have a pole at $u=0$ of order $k+1$. It is obvious that the minus sign in \eref{24} comes from the clockwise integration due to the last change of variable. Then the residue is calculated straightforwardly, one obtains
\begin{eqnarray}\label{25}
\textrm{Res}\Bigg\{\frac{\left(1-\xi u\right)^{-n}}{u^{k+1}}\Bigg\}\Bigg|_{u\to0}&=&
\frac{1}{k!}\lim_{u\to0}\frac{\rmd^k}{\rmd u^k}\left(1-\xi u\right)^{-n}\nonumber\\
&=&\frac{(n)_k}{k!}\,\xi^k,
\end{eqnarray}
where $(n)_k=n(n+1)\cdots(n+k-1)=\Gamma(n+k)/\Gamma(n)$ are the often-used Pochhammer symbol \cite{13}. This identity proves \eref{23} and thus our assertion about functions $f_{p,q}(z)$ and $g_{p,q}(z)$.
\end{proof}
\indent The reader should notice that \eref{20} can be rewritten in the Cauchy integral theorem and identifying this latter to \eref{19} leads to the identity
\begin{eqnarray}\label{26}
\frac{\lambda^n}{(n-1)!}\frac{\rmd^{n-1}}{\rmd\xi^{n-1}}\,\xi^n g_{p,q}(\xi)
=\mathcal{N}_{p,q}\oint_C z^{p-1}\left(1-z\right)^{q-1}\rme^{\lambda\xi z}\,\rmd z.
\end{eqnarray}
\indent Substituting the second summation of \eref{22} into \eref{26} and performing the $(n-1)^\textrm{th}$-derivative and expanding $\exp\left[\lambda\xi z\right]$ in the development of the Taylor series, we obtain
\begin{eqnarray}\label{27}
f_{p,q}(\xi)&=&\lambda^n\sum_{k=0}^\infty\frac{\Gamma(p+k+1)}{\Gamma(n)\Gamma(k+2)}\,c_k\,\xi^{k+1}\nonumber\\
&=&\mathcal{N}_{p,q}\sum_{k=0}^\infty\lambda^{k+1}\left(\oint_C z^{p+k}\left(1-z\right)^{q-1}\rmd z\right)\frac{\xi^{k+1}}{\Gamma(k+2)},
\end{eqnarray}
where $\Gamma$ refers to the Euler gamma function. The coefficient into the first summation in \eref{27} yields a two-term recursion relation identifying $b_{k+1}$ of \eref{22} to $c_k$, i.e.
\begin{eqnarray}\label{28}
b_{k+1}=\lambda^n\frac{\Gamma(p+k+1)}{\Gamma(n)\Gamma(k+2)}\,c_k,
\end{eqnarray}
while the comparison between the two summations leads to evaluate $\mathcal{N}_{p,q}$.\\
\indent Indeed since now $z\in\left[0,1\right]$, then the integral calculation by means of Euler's beta function is straightforward which yields
\begin{eqnarray}\label{29}
\mathcal{N}_{p,q}=\lambda^{-k}\,\frac{\Gamma(k+p+q)}{\Gamma(q)}\frac{\Gamma(k+1)}{\Gamma(k+q)}\,b_k.
\end{eqnarray}
\indent Substituting \eref{29} into the first summation of \eref{22} and by introducing the beta function, $B(p,q)$, inside the summation, we get the solution of our problem which is given by
\begin{eqnarray}\label{30}
f_{p,q}(\xi)&=&\sum_{k=0}^\infty b_{k}\xi^k\nonumber\\
&=&\mathcal{N}_{p,q}B(p,q)\sum_{k=0}^\infty\frac{(p)_k}{(p+q)_k}\frac{(\lambda\,\xi)^k}{k!}\nonumber\\
&=&\mathcal{N}_{p,q}B(p,q)\,_1F_1(p\:;p+q\:;\lambda\,\xi),
\end{eqnarray}
where $_1F_1(p\:;p+q\:;\lambda\xi)$ is the confluent hypergeometric function \cite{13}. Note from \eref{15} that $p+q=2\kappa+1$ with $\kappa=\epsilon/\alpha$.\\
\indent However it is well-known that if $p$ equals zero or a negative integer, the series in \eref{30} terminates and the confluent hypergeometric function becomes a simple polynomial. We limit ourselves here to the second case, i.e. $p=-n$. For this reason, it is convenient to redefine the confluent hypergeometric function in terms of the associated Laguerre polynomials $_1F_1(-n\:;2\kappa+1\:;\lambda\xi)=n!\,\Gamma(2\kappa+1)/\Gamma(n+2\kappa+1)\,
L_n^{(2\kappa)}(\lambda\,\xi)$, and substituting \eref{18} and \eref{30} into \eref{8} one gets
\begin{eqnarray}\label{31}
\R_{n,l}(\xi)=\mathcal{C}_{p,q}\,\xi^\kappa\,\rme^{-\frac{\lambda}{2}\,\xi}L_n^{(2\kappa)}(\lambda\,\xi),
\end{eqnarray}
where $\mathcal{C}_{p,q}=\mathcal{N}_{p,q}\,B(p,q)\,n!\,\Gamma(2\kappa+1)/\Gamma(n+2\kappa+1)$ is the correct normalization constant to be determined.\\
\indent However from the first equation of \eref{15}, i.e.
\begin{eqnarray}\label{32}
p\equiv-n=\frac{\epsilon}{\alpha}+\frac{1}{2}-\frac{\beta_1^2}{2\,\alpha\beta_2},
\end{eqnarray}
we read off the bound-state energy levels where $n$ is the vibrational quantum number.\\
\indent Indeed inserting the parameter $\epsilon$ as defined in \eref{7} into \eref{32}, the corresponding energy eigenvalues are then given by
\begin{eqnarray}\label{33}
E_{n,l}=-\frac{\hbar^2}{8mr_0^2}\left[\frac{\beta_1^2}{\beta_2}-(2n+1)\alpha\right]^2+\gamma a_0.
\end{eqnarray}
\indent The next step is devoted to evaluate, with the help of \eref{31}, the correct normalization constant $\mathcal{C}_{p,q}$ using the normalized condition
\begin{eqnarray}\label{34}
1&=&\int_{-\infty}^{+\infty}\R_{n,l}^\ast(x)\,\R_{n,l}(x)\,\rmd x\nonumber\\
&=&\int_0^{+\infty}\R_{n,l}^\ast(\xi)\,\frac{1}{\alpha\xi}\,\R_{n,l}(\xi)\,\rmd\xi.
\end{eqnarray}
\indent This can be further written as
\begin{eqnarray}\label{35}
\mathcal{C}^2_{p,q}\int_0^{+\infty}\phi^{2\kappa-1}\,\rme^{-\phi}
  \left[L_n^{(2\kappa)}(\phi)\right]^2\,\rmd\phi=\alpha\lambda^{2\kappa},
\end{eqnarray}
where $\phi=\alpha\,\xi$ and from which we obtain
\begin{eqnarray}\label{36}
\mathcal{C}_{p,q}=\left(\frac{2\beta_2}{\alpha}\right)^{\epsilon/\alpha}
\sqrt{\frac{2\,\epsilon\,\Gamma(n+1)}{\Gamma(\frac{2\epsilon}{\alpha}+n+1)}},
\end{eqnarray}
and the radial wave-function for the rotating Morse potential is given by
\begin{eqnarray}\label{37}
\R_{n,l}(\xi)=\left(\frac{2\beta_2}{\alpha}\right)^{\epsilon/\alpha}
\sqrt{\frac{2\,\epsilon\,\Gamma(n+1)}{\Gamma(\frac{2\epsilon}{\alpha}+n+1)}}
\,\xi^\kappa\,\rme^{-\frac{\lambda}{2}\,\xi}L_n^{(2\kappa)}(\lambda\,\xi).
\end{eqnarray}
\indent In the derivation of \eref{36} the correct normalization constant $\mathcal{C}_{p,q}$ can be derived from the general expression \cite{14}
\begin{eqnarray}\label{38}
\int_0^{+\infty}\xi^{\mu+\nu}\rme^{-\xi}\left[L_n^{(\mu)}(\xi)\right]^2\,\rmd\xi=
\frac{\Gamma(\mu+n+1)}{\Gamma(n+1)}\nonumber\\
\times\sum_{k=0}^{n}(-1)^k\frac{\Gamma(n-k-\nu)}{\Gamma(-k-\nu)\Gamma(n-k+1)}
\frac{\Gamma(\mu+\nu+k+1)}{\Gamma(k+1)\Gamma(n+k+1)},
\end{eqnarray}
with $\Re(\mu+\nu+1)>0$. To prove it, write one $L_n^{(\mu)}(\xi)$ in \eref{38} as the power series; i.e. $L_n^{(\mu)}(\xi)=\sum_{k=0}^n (-1)^k\frac{(n+\mu)!}{(n-k)!(k+\mu)!}\frac{\xi^k}{\Gamma(k+1)}$ and then use Eq.(\textbf{7.414}\,11) of Ref. \cite{15}.

%%%%%%%%%%%%%%%%%%%%%%%%%%%%%%%%%
\section{Numerical calculations}%
%%%%%%%%%%%%%%%%%%%%%%%%%%%%%%%%%

\indent In tables 1-4, respectively, we give some numerical applications for four (4) diatomic molecules, e.g. the H$_2$, CO, HCl and LiH using \eref{33} and then compare with the findings of SUSY, Variational, NU, SNE, MSNE, AIM and EQR. It is found that the results obtained by our approach agree with those cited above. The well-known values of relevant potential parameters are taken from Ref. \cite{9}.\\[0.0cm]

%%%%%%%%%%%%%
%  Table 1  %
%  -------  %
%%%%%%%%%%%%%

\begin{table}[h]
\caption{\label{table 1} Energy eigenvalues (in eV) for the H$_2$ diatomic molecule.}
\begin{indented}
\lineup
\item[]\begin{tabular}{@{}*{7}{l}}
\br
$n$&$\0l$&ICR (present work)&SUSY \cite{4}&AIM \cite{9}&Variational \cite{5}&SNE \cite{7}\cr
\mr
0&\00 &$-$4.47600&$-$4.47601&$-$4.47601&$-$4.4758&$-$4.4749\cr
 &\05 &$-$4.25879&$-$4.25880&$-$4.25880&$-$4.2563&$-$4.2590\cr
 &10  &$-$3.72192&$-$3.72193&$-$3.72193&$-$3.7187&$-$3.7247\cr
5&\00 &$-$2.22051&$-$2.22051&$-$2.22052&\0\0\0---&$-$2.2038\cr
 &\05 &$-$2.04353&$-$2.04353&$-$2.04355&\0\0\0---&$-$2.0525\cr
 &10  &$-$1.60389&$-$1.60389&$-$1.60391&\0\0\0---&$-$1.6526\cr
7&\00 &$-$1.53743&$-$1.53743&$-$1.53744&\0\0\0---&$-$1.5168\cr
 &\05 &$-$1.37654&$-$1.37654&$-$1.37656&\0\0\0---&$-$1.3887\cr
 &10  &$-$0.97579&$-$0.97578&$-$0.97581&\0\0\0---&$-$1.0499\cr
\br
\end{tabular}
\end{indented}
\end{table}

%%%%%%%%%%%%%
%  Table 2  %
%  -------  %
%%%%%%%%%%%%%

\begin{table}[h]
\caption{\label{table 2} Energy eigenvalues (in eV) for the CO diatomic molecule.}
\begin{indented}
\lineup
\item[]\begin{tabular}{@{}*{7}{l}}
\br
$n$&$\0l$&\0ICR (present work)&\0EQR \cite{10}&Variational \cite{5}&\0NU \cite{6}&\0SNE \cite{7}\cr
\mr
0&\00 &$-$11.0911  &$-$11.0915  &$-$11.093&$-$11.091&$-$11.091\cr
 &\05 &$-$11.0839  &$-$11.0844  &$-$11.084&$-$11.084&$-$11.084\cr
 &10  &$-$11.0649  &$-$11.0653  &$-$11.0653&$-$11.065&$-$11.065\cr
5&\00 &\0$-$9.79475&\0$-$9.79519&\0\0\0---&\0$-$9.795&\0$-$9.788\cr
 &\05 &\0$-$9.78791&\0$-$9.78835&\0\0\0---&\0$-$9.788&\0$-$9.782\cr
 &10  &\0$-$9.76967&\0$-$9.77011&\0\0\0---&\0$-$9.769&\0$-$9.765\cr
7&\00 &\0$-$9.29876&\0$-$9.2992 &\0\0\0---&\0$-$9.299&\0$-$9.286\cr
 &\05 &\0$-$9.29204&\0$-$9.29248&\0\0\0---&\0$-$9.292&\0$-$9.281\cr
 &10  &\0$-$9.27413&\0$-$9.27457&\0\0\0---&\0$-$9.274&\0$-$9.265\cr
\br
\end{tabular}
\end{indented}
\end{table}

%%%%%%%%%%%%%
%  Table 3  %
%  -------  %
%%%%%%%%%%%%%

\begin{table}[h]
\caption{\label{table 3} Energy eigenvalues (in eV) for the HCl diatomic molecule.}
\begin{indented}
\lineup
\item[]\begin{tabular}{@{}*{7}{l}}
\br
$n$&$\0l$&ICR (present work)&SUSY \cite{4}&AIM \cite{9}&Variational \cite{5}&MSNE \cite{8}\cr
\mr
0&\00 &$-$4.43537&$-$4.43556&$-$4.4356&$-$4.4360&$-$4.4355\cr
 &\05 &$-$4.39663&$-$4.39681&$-$4.3968&$-$4.3971&$-$4.3968\cr
 &10  &$-$4.29389&$-$4.29408&$-$4.2941&$-$4.2940&$-$4.2940\cr
5&\00 &$-$2.80490&$-$2.80508&$-$2.8051&\0\0\0---&$-$2.8046\cr
 &\05 &$-$2.77193&$-$2.77211&$-$2.7721&\0\0\0---&$-$2.7718\cr
 &10  &$-$2.68455&$-$2.68473&$-$2.6847&\0\0\0---&$-$2.6850\cr
7&\00 &$-$2.25686&$-$2.25703&$-$2.2570&\0\0\0---&$-$2.2565\cr
 &\05 &$-$2.22619&$-$2.22636&$-$2.2263&\0\0\0---&$-$2.2262\cr
 &10  &$-$2.14497&$-$2.14513&$-$2.1451&\0\0\0---&$-$2.1461\cr
\br
\end{tabular}
\end{indented}
\end{table}

%%%%%%%%%%%%%
%  Table 4  %
%  -------  %
%%%%%%%%%%%%%

\begin{table}[h]
\caption{\label{table 4} Energy eigenvalues (in eV) for the LiH diatomic molecule.}
\begin{indented}
\lineup
\item[]\begin{tabular}{@{}*{7}{l}}
\br
$n$&$\0l$&ICR (present work)&EQR \cite{10}&AIM \cite{9}&MSNE \cite{8}&NU \cite{6}\cr
\mr
0&\00 &$-$2.42876&$-$2.42886&$-$2.4289&$-$2.4280&$-$2.4287\cr
 &\05 &$-$2.40123&$-$2.40133&$-$2.4013&$-$2.4000&$-$2.4012\cr
 &10  &$-$2.32873&$-$2.32883&$-$2.3288&$-$2.3261&$-$2.3287\cr
5&\00 &$-$1.64763&$-$1.64772&$-$1.6477&$-$1.6402&$-$1.6476\cr
 &\05 &$-$1.62368&$-$1.62377&$-$1.6238&$-$1.6160&$-$1.6236\cr
 &10  &$-$1.56065&$-$1.56074&$-$1.5607&$-$1.5525&$-$1.5606\cr
7&\00 &$-$1.37748&$-$1.37757&$-$1.3776&$-$1.3682&$-$1.3774\cr
 &\05 &$-$1.35496&$-$1.35505&$-$1.3550&$-$1.3456&$-$1.3549\cr
 &10  &$-$1.29572&$-$1.29581&$-$1.2958&$-$1.2865&$-$1.2957\cr
\br
\end{tabular}
\end{indented}
\end{table}

%%%%%%%%%%%%%%%%%%%%%
\section{Conclusion}%
%%%%%%%%%%%%%%%%%%%%%

\indent We have investigated a method of inverse contour representation in the context of a contour Laplace transforms in order to solve the bound-states problem for the one-dimensional Schr\"odinger equation related to the rotating Morse potential. The main purpose of the approach developed here is to investigate how this method works. In general the determination of the inverse contour transform is the main problem using the integral transformation, unless if it is represented by a known function like here, e.g. the Euler beta function. In fact using Cauchy's integral formula, the beta function may be used to advantage in the evaluation of the inverse contour representation which leads to the application of the calculus of residues. It is perhaps worth noting that this approach was successful and relatively easy, because there is not a specific choice for the contour of integration $C$ in which a lot of the possible types of contours are unlimited. We have shown that the inverse contour representation opens up the way for another and new derivation of the bound-states for the rotating Morse potential.\\
\indent Finally it would be interesting to complete our approach for different types of potentials belonging to the hypergeometric differential equation, e.g. P\"oschl-Teller and others mentioned in \cite{16}.

\end{document}